\documentclass[journal=nalefd,manuscript=article]{achemso}

\usepackage[version=3]{mhchem} 
\usepackage[english]{babel}
\usepackage{units}
\usepackage{amssymb}
\usepackage{gensymb}
\usepackage{xcolor}
\usepackage{setspace}

\let\oldphi\phi \let\phi\varphi \let\varphi\oldphi 

\author{\fontfamily{ptm}\selectfont Lars Sjöström}
\email{sjolars@chalmers.se}
\author{Prasanna Rout}
\affiliation[Chalmers University of Technology]{Department of Microtechnology and Nanoscience, Chalmers University of Technology, SE-41296 Gothenburg, Sweden}
\author{Shahid Sattar}
\affiliation[Linnaeus University]{Department of Mathematics and Physics, Linnaeus University, SE-39231 Kalmar, Sweden}
\author{Alexander Tyner}
\affiliation[Nordita]{Nordita, KTH Royal Institute of Technology and Stockholm University, Hannes Alfvéns väg 12, SE-10691 Stockholm, Sweden}
\alsoaffiliation{Department of Physics, University of Connecticut, Storrs, Connecticut 06269, USA}
\author{Maurice E. Bal}
\affiliation[HFML Nijmegen]{High Field Magnet Laboratory (HFML -- EMFL), Radboud University, 6525 ED Nijmegen, The Netherlands}
\author{Ankit Khola}
\affiliation[Chalmers University of Technology]{Department of Microtechnology and Nanoscience, Chalmers University of Technology, SE-41296 Gothenburg, Sweden}
\author{Elias Rasmussen}
\affiliation[Chalmers University of Technology]{Department of Microtechnology and Nanoscience, Chalmers University of Technology, SE-41296 Gothenburg, Sweden}
\author{Khadiza Ali}
\affiliation[Chalmers University of Technology]{Department of Microtechnology and Nanoscience, Chalmers University of Technology, SE-41296 Gothenburg, Sweden}
\author{Arumugum Thamizhavel}
\affiliation[TIFR]{Department of Condensed Matter Physics and Materials Science, Tata Institute of Fundamental Research, Mumbai 400005, India}
\author{Uli Zeitler}
\affiliation[HFML Nijmegen]{High Field Magnet Laboratory (HFML -- EMFL), Radboud University, 6525 ED Nijmegen, The Netherlands}
\author{Carlo M. Canali}
\affiliation[Linnaeus University]{Department of Mathematics and Physics, Linnaeus University, SE-39231 Kalmar, Sweden}
\author{Saroj P. Dash}
\email{saroj.dash@chalmers.se}
\affiliation[Chalmers University of Technology]{Department of Microtechnology and Nanoscience, Chalmers University of Technology, SE-41296 Gothenburg, Sweden}
\alsoaffiliation{Wallenberg Initiative Materials Science for Sustainability, Chalmers University of Technology, SE-41296 Gothenburg, Sweden}
\alsoaffiliation{Graphene Center, Chalmers University of Technology, SE-41296 Gothenburg, Sweden}

\title{\singlespacing{\fontfamily{ptm}\selectfont Revealing quantum metric multipoles in magnetic topological insulator \ce{MnBi2Te4}}}


\begin{document}



\newpage
\begin{abstract}
\singlespacing

Nonlinear electronic transport has emerged as a powerful probe of the quantum geometry in topological quantum materials where the band topology and broken symmetries facilitate power-law current-voltage responses beyond Ohm's law. While nonlinear transport of the second and third orders has been studied in several quantum materials, higher-order transport has so far mainly remained experimentally inaccessible, leaving more detailed features of the quantum geometry unexplored. Here, we observe higher-order nonlinear electronic transport up to the seventh harmonic order in multilayer magnetic topological insulator \ce{MnBi2Te4}. We find an even-odd behavior where the odd-order nonlinear transport components dominate while the even-order ones are suppressed. Temperature- and magnetic-field-dependent measurements show a strong correlation between the nonlinear transport and the magnetic phases of \ce{MnBi2Te4}. Through scaling analysis and theoretical calculations, quantum metric multipoles and nonlinear Drude conductivities are identified as the microscopic origins of the nonlinear transport.

\end{abstract}

\singlespacing

\section{Introduction}

Distinctly different from ohmic conduction, nonlinear electronic transport with power-law current-voltage relationships can arise due to the quantum geometry in materials with a rich band topology and low symmetry.\cite{Suarez-Rodriguez2025NonlinearSystems,Zhu2025MagneticTransports,Gong2025NonlinearMetric,Liu2022BerryEffect} Nonlinear transport measurements have therefore recently emerged as an important probe of the quantum geometry (consisting of its real and imaginary parts: the quantum metric and the Berry curvature, respectively) as well as of scattering effects such as side jump and skew scattering.\cite{Suarez-Rodriguez2025NonlinearSystems,Verma2026QuantumMaterials,Torma2022SuperconductivitySystems,Liu2021IntrinsicAntiferromagnets,Wang2021IntrinsicCuMnAs,Liu2022BerryEffect,Ma2023AnomalousAntiferromagnets,Fang2024QuantumAltermagnets,Zhu2025MagneticTransports} While the Berry curvature is well-established as a cornerstone in topological physics,\cite{Xiao2010BerryProperties,Chang2008BerryFields}  the quantum metric is relatively unexplored despite having an important influence on nonlinear transport, superconductivity and flat-band physics.\cite{Kim2025DirectSolids,Verma2026QuantumMaterials,Torma2022SuperconductivitySystems,Liang2017BandWeight,Rhim2020QuantumBands,Gao2023QuantumHeterostructure,Wang2023Quantum-metric-inducedAntiferromagnet,Li2024QuantumMnBi2Te4,Han2024Room-temperatureAntiferromagnet,Kang2025MeasurementsSolids,Jin2026ExperimentalMetric,Diez-Carlon2025ProbingJunctions}

\begin{figure}[ht!]
    \centering
    \includegraphics[width=\textwidth]{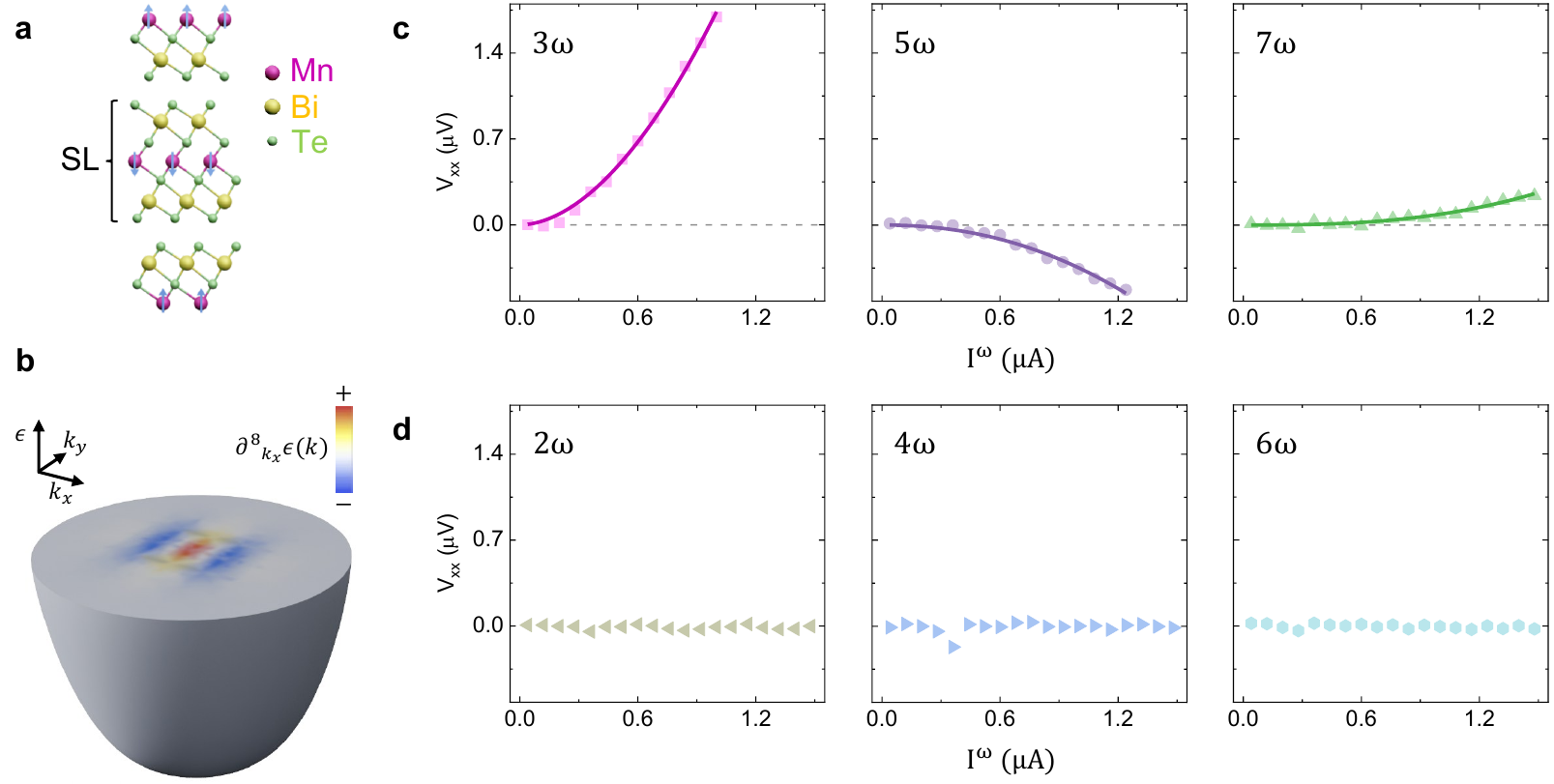}
    \caption{{\footnotesize \textbf{Higher-harmonic nonlinear transport in multilayer \ce{MnBi2Te4}.} \textbf{(a)} Side-view schematic of the crystal structure of \ce{MnBi2Te4}, showcasing the antiferromagnetic ordering of adjacent septuple layers (SLs) with an out-of-plane magnetization. The blue arrows indicate the spin direction of the respective Mn atoms. \textbf{(b)} Schematic of \ce{MnBi2Te4} band diagram with a quantum metric septupole as a color plot, which corresponds to the seventh-order nonlinear conductivity generated from the quantum metric. \textbf{(c)} Nonlinear $IV$ characteristics for the third ($3\omega$), fifth ($5\omega$) and seventh ($7\omega$) harmonic orders with power-law fits as guides to the eye (solid lines). \textbf{(d)} Nonlinear $IV$ characteristics for the second ($2\omega$), fourth ($4\omega$) and sixth ($6\omega$) harmonic orders, showing no signals. The measurements were performed in Device 1  at $T=\unit[2]{K}$.}}
    \label{fig:Higher_harmonic_IV}
\end{figure}

The magnetic topological insulator \ce{MnBi2Te4} (MBT) has proven to be an ideal platform for investigating nonlinear transport originating from the quantum geometry, having an A-type antiferromagnetic (AFM) ordering (illustrated in Figure \ref{fig:Higher_harmonic_IV}a) along with a rich topological band structure which hosts a range of exotic effects, including the quantum anomalous Hall and axion insulator states.\cite{Li2019IntrinsicMaterials,Li2024ProgressMnBi2Te4,Deng2020QuantumMnBi2Te4,Lian2025AntiferromagneticFlops,Liu2020RobustInsulator,Qiu2025ObservationMnBi2Te4} Recently, second-order nonlinear transport generated by a quantum metric dipole was demonstrated in thin even-layer MBT with a broken inversion symmetry,\cite{Gao2023QuantumHeterostructure,Wang2023Quantum-metric-inducedAntiferromagnet} whereas third-order transport due to quantum metric and Berry curvature quadrupoles was observed in bulk MBT with maintained inversion symmetry.\cite{Li2024QuantumMnBi2Te4} However, while higher-order transport has been theoretically predicted to reveal exotic higher-order multipoles in the quantum geometry,\cite{Zhang2023Higher-orderMultipoles} experimental harmonic measurements in all studied materials have so far almost exclusively been limited to the second and third harmonic orders.\cite{Ma2019ObservationConditions,Kang2019NonlinearWTe2,Kumar2021Room-temperatureTaIrTe4,Lai2021Third-orderTensor,Wang2022Room-temperatureTaIrTe4,Ma2022GrowthEffect,Lee2024Spin-orbit-splitting-drivenNbIrTe4,Jiang2025ProbingEffect,Xi2025TerahertzTaIrTe4}

Here, we demonstrate higher-order nonlinear transport in MBT up to the seventh harmonic order. An even-odd behavior is observed in the harmonic order, where the odd-order harmonic signals are finite while the even-order ones are significantly suppressed. The nonlinear transport signals, measured as a function of temperature and magnetic fields, are correlated with the magnetic phase transitions in MBT. Finally, the microscopic origin of the nonlinear transport is investigated through scaling-law analysis and theoretical calculations, which indicate contributions from quantum metric multipoles (illustrated in Figure \ref{fig:Higher_harmonic_IV}b) as well as higher-order Drude-like nonlinear conductivities. These results pave the path towards a deeper understanding of complex quantum geometric properties such as higher-order multipoles in exotic topological quantum materials through nonlinear transport measurements.

\section{Results and discussion}

\subsection{Observation of higher-harmonic transport signals}

\begin{figure}[ht!]
    \centering
    \includegraphics[width=0.6\textwidth]{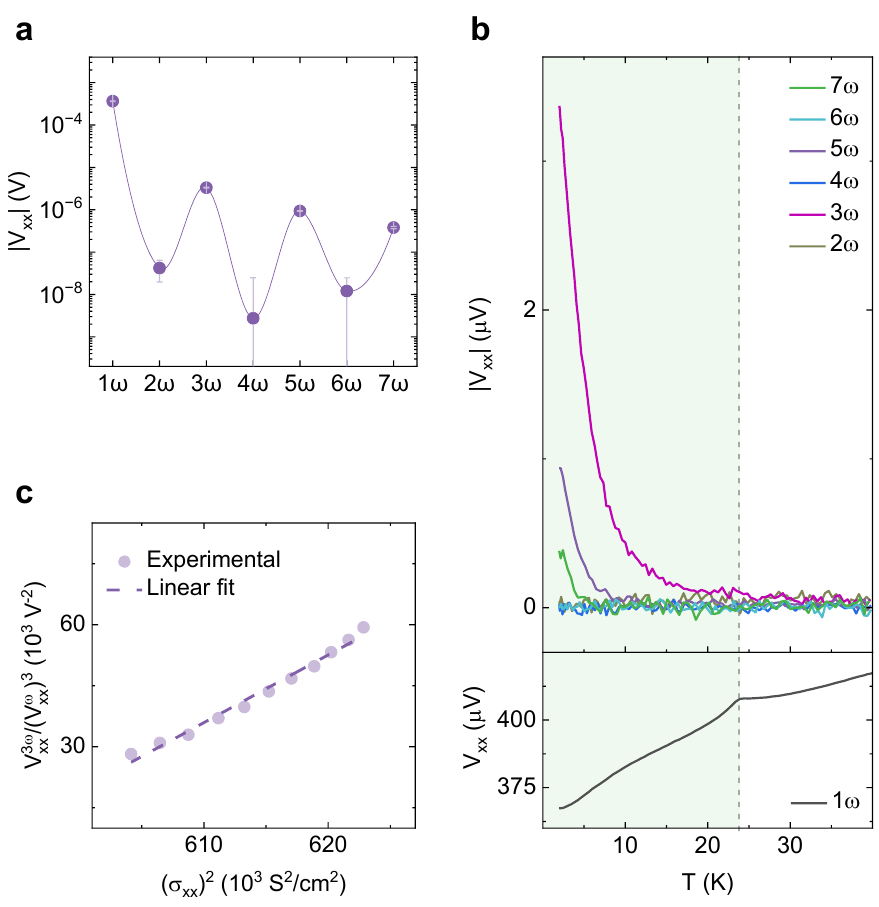}
    \caption{{\footnotesize \textbf{Even-odd behavior and temperature-dependent higher-harmonic nonlinear transport.} \textbf{(a)} The nonlinear voltage signals for different harmonic orders at $T = \unit[2]{K}$, showing an even-odd behavior as well as a decreasing amplitude for increasing harmonic orders. \textbf{(b)} Nonlinear voltages $|V_{xx}|$ of the second ($2\omega$) to the seventh ($7\omega$) harmonic order as a function of temperature. The first-harmonic signal is shown in the bottom panel for reference. The Néel temperature $T_N=\unit[23.8]{K}$ is indicated with a dashed line. \textbf{(c)} The third-harmonic voltage normalized by the cube of the first-harmonic voltage plotted against the square of the longitudinal conductivity for low temperatures $T=\unit[2.6-5.3]{K}$. The data follows a linear fit (dashed line) according to the scaling formula in Equation \eqref{eq:ScalingLaw}. The measurements were performed in Device 1 with $I^\omega = \unit[2]{\mu A}$.}}
    \label{fig:HigherHarmonic_Temp}
\end{figure}

The nonlinear transport was probed through higher-harmonic measurements in Hall-bar devices which were fabricated from exfoliated multilayers of MBT ($\unit[60-220]{nm}$) and covered by \ce{Al2O3} to prevent oxidation. By measuring at a higher frequency $n\omega$ with respect to the applied AC current $I^{\omega}$ with frequency $\omega$ using a lock-in technique, the $n$\textsuperscript{th}-harmonic component of the nonlinear voltage response could be extracted. The third- ($3\omega$), fifth- ($5\omega$) and seventh-harmonic ($7\omega$) longitudinal voltage $V_{xx}$ responses are presented in Figure \ref{fig:Higher_harmonic_IV}c, clearly displaying nonlinear $IV$ characteristics. The linear first-harmonic voltage $V_{xx}^\omega$ is shown in Supplementary Figure S1 for reference. In contrast, the even-harmonic (\emph{i.e.} second- [$2\omega$], fourth- [$4\omega$] and sixth-harmonic [$6\omega$]) signals are significantly suppressed (shown in Figure \ref{fig:Higher_harmonic_IV}d and in Supplementary Figure S2). Additionally, an overall trend of rapidly decreasing signal amplitudes for increasing harmonic orders is observed. The third-harmonic signal clearly dominates over the fifth-harmonic one, which in turn is significantly larger than the seventh-harmonic signal. Indeed, this is generally expected since lower-harmonic signals tend to dominate in transport measurements unless they are suppressed by symmetry.\cite{Zhang2023Higher-orderMultipoles} The amplitude of these voltage signals is plotted for different harmonic orders in Figure \ref{fig:HigherHarmonic_Temp}a, which clearly summarizes both the even-odd behavior and the reduction in amplitudes with increasing harmonic orders.

Before investigating the higher-harmonic signals further, various spurious effects that can lead to nonlinear transport were addressed. Thermoelectric effects are expected to persist above the Néel temperature of MBT and grow with increasing temperatures,\cite{Yan2019CrystalMnBi2Te4,Chai2024ThermoelectricMnBi2Te4} very differently from our observed signals below. Additionally, the small current dependence of the magnetic phase transitions indicates only minor overall heating in our devices, and the normalized third-harmonic signal has an opposite trend with an increasing current compared to what is expected from thermoelectric contributions.\cite{Li2024QuantumMnBi2Te4} Contributions from capacitive coupling are ruled out due to the absence of a frequency dependence of the higher-harmonic signals.\cite{Kang2019NonlinearWTe2} Ohmic contact behavior and highly symmetric shapes of the MBT flakes indicate no effects from the device geometry or contact interfaces. Mixing effects from lower- to higher-order harmonic signals are excluded analytically, and the measured lock-in phase of the higher-harmonic signals confirm the absence of artificial signal contributions. In this way, spurious effects could be ruled out as main contributors to the observed nonlinear transport signals, which is discussed in further detail in Supplementary Note 1.

\subsection{Temperature dependence and conductivity scaling analysis}

The temperature dependence of the harmonic signals is depicted in Figure \ref{fig:HigherHarmonic_Temp}b. With decreasing temperatures, MBT transitions from a paramagnetic (PM) to an AFM phase. This transition can be seen as a kink in the first-harmonic signal at the Néel temperature $T_N=\unit[23.8]{K}$, which is a typical value for multilayer MBT.\cite{Deng2020QuantumMnBi2Te4,Liu2020RobustInsulator,Zhang2022ControlledMnBi2Te4} Below $T_N$, the third-harmonic signal increases sharply as the temperature decreases. Both the fifth- and the seventh-harmonic voltages show a similar behavior, although with significantly lower amplitudes in accordance with Figure \ref{fig:HigherHarmonic_Temp}a. The temperature dependence thus suggests a strong correlation between the nonlinear transport and the magnetization of MBT. In contrast, the even-order harmonic voltages vanish at all temperatures for the same applied current.

To find the underlying mechanism of the nonlinear transport, conductivity scaling analysis of the third-harmonic signal was performed, where the quantum metric and Drude-like (\emph{i.e.} extrinsic scattering-governed\cite{Suarez-Rodriguez2025NonlinearSystems,Gong2025NonlinearMetric}) contributions could be separated due to their different dependence on the scattering time. In general, nonlinear transport can also be generated by Berry curvature multipoles, but this is forbidden for longitudinal voltages by the $C_3$ crystal symmetry of MBT.\cite{Li2024QuantumMnBi2Te4,Suarez-Rodriguez2025NonlinearSystems,Liu2021IntrinsicAntiferromagnets,Wang2021IntrinsicCuMnAs} Figure \ref{fig:HigherHarmonic_Temp}c shows the normalized third-harmonic voltage $V^{3\omega}_{xx}/\left(V^\omega_{xx}\right)^3$ \emph{versus} the square of the longitudinal conductivity $\sigma_{xx}$.
The Drude-like and quantum metric components of the nonlinear transport correspond to the slope $\alpha$ and the intercept $\beta$, respectively, of the scaling formula\cite{Li2024QuantumMnBi2Te4}
\begin{equation}
    \frac{V_{xx}^{3\omega}}{\left(V_{xx}^\omega\right)^3} = \alpha {\sigma_{xx}}^2 + \beta .
    \label{eq:ScalingLaw}
\end{equation}
The fit to the data indicates that both effects contribute almost equally to the nonlinear conductivity (see Supplementary Note 2 for details). Such near-equal contributions from the quantum metric and the Drude-like conductivities were also observed in additional devices, as shown in Supplementary Note 2. No scaling analysis could be performed for the fifth- and seventh-harmonic signals since conductivities from the quantum metric have not been derived beyond the third order,\cite{Liu2022BerryEffect} but contributions from quantum metric and Drude-like conductivities are expected.

\subsection{Magnetic field dependence and magnetic phase correlation}

\begin{figure}[ht!]
    \centering
    \includegraphics[width=\textwidth]{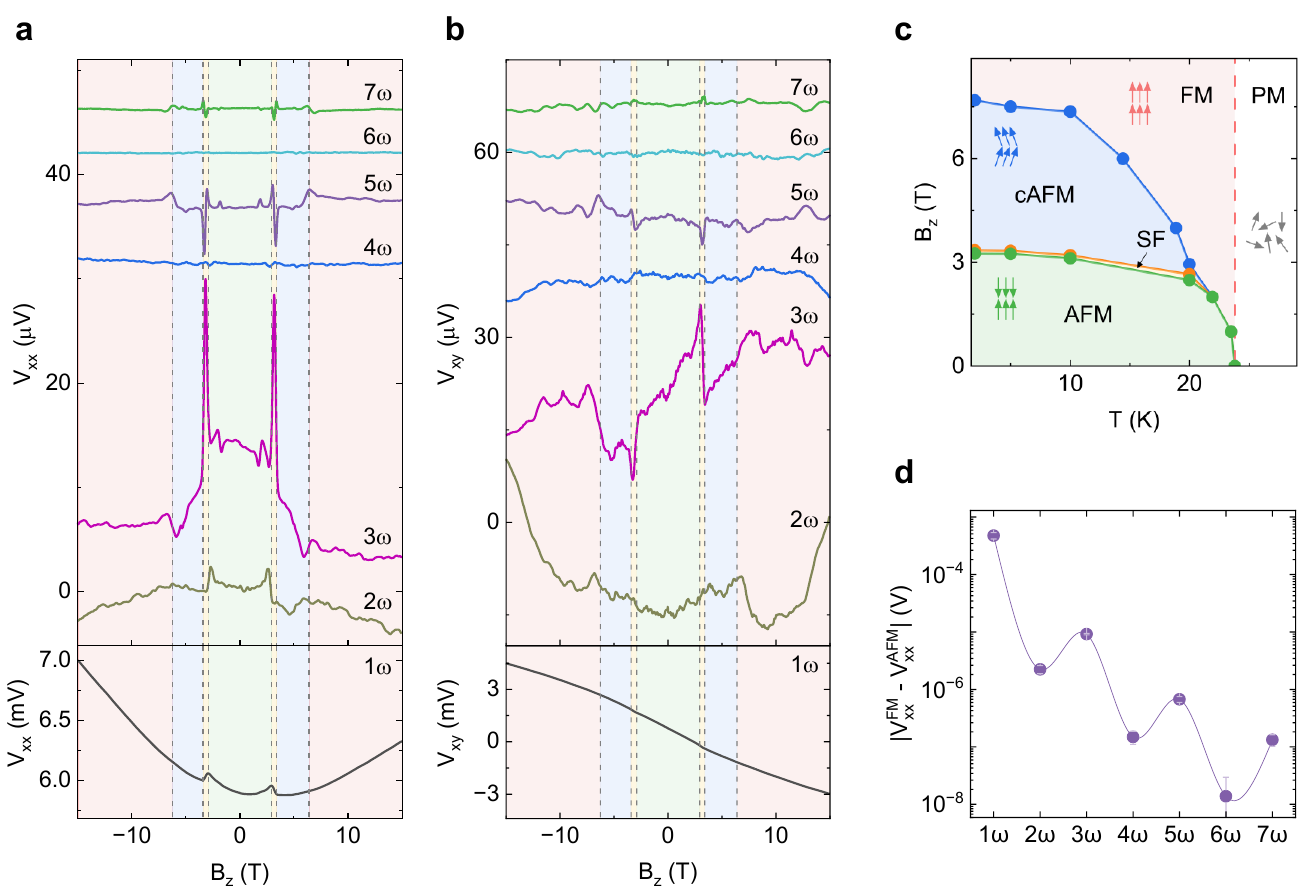}
    \caption{{\footnotesize \textbf{Magnetic-field-dependent higher-order nonlinear transport.} \textbf{(a,b)} Higher-harmonic longitudinal voltages $V_{xx}$ (a) and transverse voltages $V_{xy}$ (b) of the second ($2\omega$) to the seventh ($7\omega$) (top) and the first ($1\omega$, bottom) harmonic orders for OOP magnetic field sweeps. The measurements were performed in Device 2 with $I^\omega = \unit[80]{\mu A}$ at $T = \unit[4.3]{K}$. The data has been shifted vertically for clarity and the magnetic phase transitions are indicated with dashed lines. \textbf{(c)} A magnetic phase diagram for MBT, where the borders between the different magnetic phases have been mapped out from linear transport measurements. Antiferromagnetic (AFM), canted antiferromagnetic (cAFM), ferromagnetic (FM) and paramagnetic (PM) phases as well as a spin-flop (SF) transition region are labeled. Colored arrows indicate the magnetic ordering of MBT in each phase. The measurements were performed in Device 3 with $I^\omega = \unit[20]{\mu A}$. \textbf{(d)} Variations in the difference in $V_{xx}$ between the AFM and the FM phases from (a) for different harmonic orders, showing an even-odd behavior as well as a decreasing amplitude for increasing harmonic orders.}}
    \label{fig:HigherHarmonic_Field}
\end{figure}

Next, the evolution of the higher-harmonic voltage responses in the presence of an out-of-plane (OOP) magnetic field $B_z$ is discussed. Interestingly, all nonlinear signals up to the seventh harmonic order are discernible with significantly different amplitudes, as seen in Figure \ref{fig:HigherHarmonic_Field}a. The even-order signals are strongly suppressed whereas the lower harmonics dominate among the odd-order signals for all magnetic fields. Sharp kinks in the nonlinear voltages are observed around $B_z=\unit[\pm 3]{T}$ and $\unit[\pm 6]{T}$. To identify the origin of these kinks, the field values where they appear are compared to the magnetic phases of MBT. The magnetic phase diagram in Figure \ref{fig:HigherHarmonic_Field}c was constructed from the temperature and OOP magnetic field dependencies of the first-harmonic signal, as described in Supplementary Note 3. The four distinct magnetic phases of MBT are identified as an AFM, a canted AFM (cAFM), a ferromagnetic (FM) and a PM phase.\cite{Bac2022TopologicalCanting,Lian2025AntiferromagneticFlops,Ye2022NonreciprocalPt,Lee2019SpinMnBi2Te4} Additionally, a narrow spin-flop (SF) transition region is visible between the AFM and the cAFM phases.\cite{Bac2022TopologicalCanting,Lian2025AntiferromagneticFlops} In this way, the kinks in the higher-harmonic transport signals could be correlated with magnetic phase transitions. As the applied magnetic field increases, the AFM ground state transitions into a cAFM phase, which gives rise to the first kinks near $B_z=\unit[\pm 3]{T}$ in the higher-harmonic voltage signals. A further increase leads to the transition from the cAFM to the FM phase, which is reflected as the second kinks around $\unit[\pm 6]{T}$. Similar behavior was observed also for the transverse voltages $V_{xy}$ in Figure \ref{fig:HigherHarmonic_Field}b.

In addition to the kinks associated with magnetic phase transitions, there is a significant shift in the longitudinal nonlinear voltages in the FM phase ($V_{xx}^{FM}$) compared to those in the AFM phase ($V_{xx}^{AFM}$), as summarized in Supplementary Figure S11. This can be clearly seen in Supplementary Figure S6, where the voltage signals without vertical shifts are shown. The amplitude difference $|V_{xx}^{FM}-V_{xx}^{AFM}|$ for all harmonic orders is presented in Figure \ref{fig:HigherHarmonic_Field}d and follows similar trends as those of the zero-field harmonic voltages in Figure \ref{fig:HigherHarmonic_Temp}a, \emph{i.e.} (1) the voltage difference for the odd-order harmonic signals are orders of magnitude larger than the one for the even-order signals, and (2) the overall response decreases with increasing harmonic orders. The shifts in the nonlinear voltages appear due to changes in the nonlinear transport contributions from the quantum metric during the transition from the AFM to the FM phase, considering that the nonlinear Drude-like conductivities are independent of magnetic phase transitions in MBT.\cite{Li2024QuantumMnBi2Te4}

\subsection{Theoretical calculations}

To understand the microscopic origin of the experimental results, theoretical calculations were performed where the band structure of the multilayer MBT was modeled using a $k.p$ model, which also allows the inclusion of perturbations to consider the effects of finite magnetic fields (see Supplementary Note 4.1). First, the Drude-like contributions to the higher-order longitudinal conductivities were calculated from higher-order differentiation of the modeled dispersion relation (described in detail in Supplementary Note 4.2), as summarized qualitatively in Figure \ref{fig:HigherHarmonic_Theory}a and shown in detail in Supplementary Figure S13. Clearly, it is noted that the different-order derivatives of the dispersion relation of MBT change between being antisymmetric and symmetric with respect to $k_x$ for even and odd orders, respectively. After integration over the reciprocal space, this leads to vanishing contributions to the second-, fourth- and sixth-order conductivities while the third-, fifth- and seventh-order conductivities are nonzero. This even-odd behavior aligns well with the experimental observations. Quantitative values for the generated odd-harmonic nonlinear currents under an applied bias voltage decrease rapidly with increasing harmonic orders (shown in Supplementary Figure S14), which is similar to the experimental results in Figure \ref{fig:HigherHarmonic_Temp}a and Figure \ref{fig:HigherHarmonic_Field}d.

\begin{figure}[ht!]
    \centering
    \includegraphics[width=\textwidth]{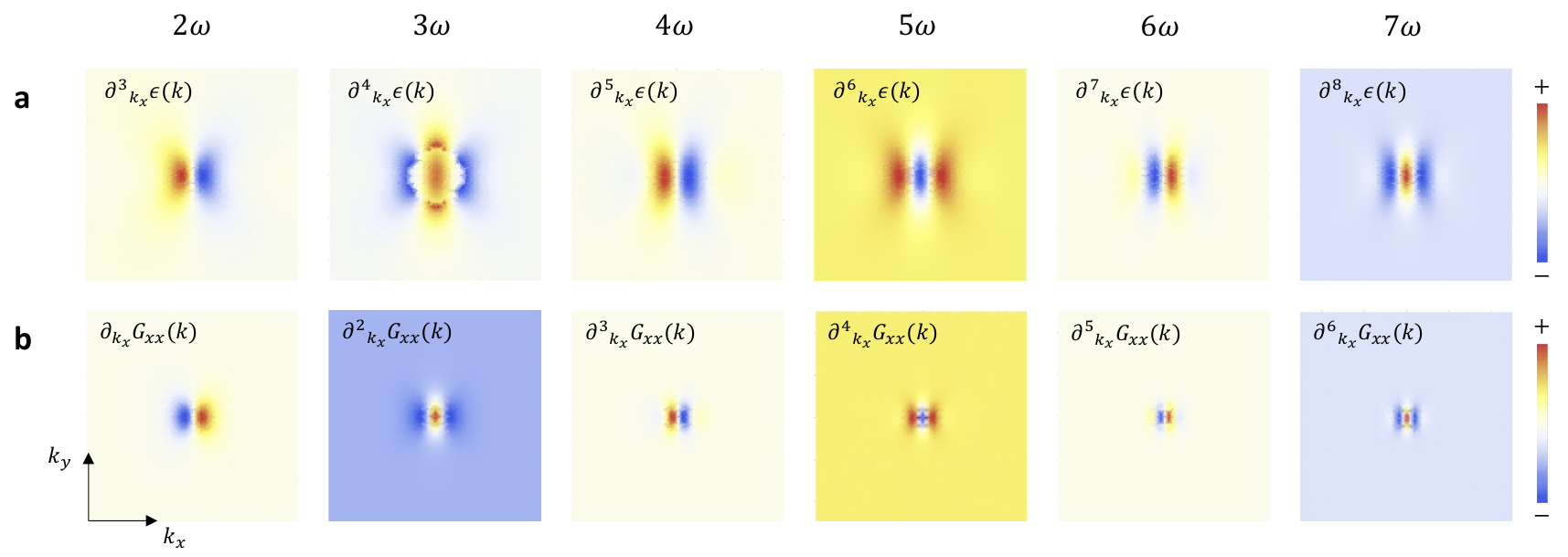}
    \caption{{\footnotesize \textbf{Theory for higher-order Drude and quantum metric contributions.} Qualitative overview of the calculated contributions to the nonlinear conductivities of second ($2\omega$) to seventh ($7\omega$) harmonic order from the Drude-like terms (a) and the quantum metric (b), respectively.}}
    \label{fig:HigherHarmonic_Theory}
\end{figure}

Next, the contributions to the higher-order longitudinal conductivities from the quantum metric were calculated from higher-order differentiation of the band-energy normalized quantum metric tensor under a finite magnetic field, as described in Supplementary Note 4.3. While the nonlinear Drude conductivity in MBT is insensitive to magnetic fields,\cite{Li2024QuantumMnBi2Te4} the quantum metric is expected to be the origin of the observed field-dependent features of the nonlinear transport. This therefore allows the quantum metric contributions to be examined in comparison to the field-insensitive Drude contributions. Interestingly, the quantum metric multipoles (summarized qualitatively in Figure \ref{fig:HigherHarmonic_Theory}b and shown in detail in Supplementary Figure S15) exhibit similar symmetries with respect to $k_x$ as the derivatives of the dispersion relation above, and integration over the reciprocal space leads to an even-odd behavior. Therefore, the quantum-metric contributions to the nonlinear transport show agreement with the even-odd behavior of the experimental results as well, similar to the Drude-like contributions, and both mechanisms can give rise to the experimentally observed nonlinear transport. The relative contributions from the Drude-like terms and the quantum metric can unfortunately not be compared directly in the theoretical calculations, since longitudinal conductivities generated by metric tensor multipoles have so far not been derived beyond the third order.\cite{Liu2022BerryEffect} Nevertheless, both mechanisms are expected to be present. Importantly, this is in agreement with the scaling analysis shown in Figure \ref{fig:HigherHarmonic_Temp}c and Supplementary Note 2, where Drude-like conductivities and the quantum metric are found to give equal contributions to the nonlinear transport.

\section*{Conclusions and outlook}

In conclusion, we have demonstrated nonlinear electronic transport up to the seventh harmonic order in multilayer MBT. An even-odd behavior with harmonic orders was found where all odd-order harmonic signals significantly dominate over the even-order ones. By correlating the harmonic transport signals with magnetic phase transitions at different temperatures and magnetic fields, the nonlinear transport was found to depend strongly on the magnetic phase of MBT. Conductivity scaling analysis and theoretical calculations indicate quantum metric multipoles and Drude-like nonlinear conductivity terms as the microscopic origin of the nonlinear transport. These findings establish higher-order nonlinear transport as a tool for the detailed exploration of exotic topological features such as quantum metric multipoles in topological quantum materials. A more comprehensive understanding of the quantum metric and its interaction with electronic transport can become a cornerstone of novel electronics and topological physics,\cite{Kim2025DirectSolids,Verma2026QuantumMaterials,Torma2022SuperconductivitySystems} and be pivotal for innovative nonlinear-transport-based technologies such as wireless rectification\cite{Kumar2021Room-temperatureTaIrTe4,Cheng2024GiantTellurium,Xi2025TerahertzTaIrTe4} and nonvolatile memories.\cite{Godinho2018ElectricallyAntiferromagnet,Nishijima2023FerroicTemperature,Jo2025AnomalousCrSBr}

\section*{Methods}
\emph{Device fabrication:} The \ce{MnBi2Te4} flakes were mechanically exfoliated and transferred onto Si/\ce{SiO2} chips inside a glovebox in \ce{N2} atmosphere. Electrodes were made using electron beam lithography (EBL) and thin film deposition of $\unit[20]{nm}$ Ti/$\unit[200]{nm}$ Au \emph{via} electron beam evaporation followed by lift-off. The entire chips were then covered by $\unit[3]{nm}$ \ce{Al2O3} using electron beam evaporation to prevent oxidation of the \ce{MnBi2Te4}. Finally, the \ce{MnBi2Te4} in Device 1, 2 and 4 were etched into Hall bars using EBL patterning and Ar milling. A representative device is shown in Supplementary Figure S16, and atomic force microscopy images are presented in Supplementary Figure S17, where the \ce{MnBi2Te4} thickness range is determined to approximately $\unit[60-220]{nm}$.

\emph{Electrical characterization and measurements:} The transport measurements in Device 2 were performed in a high-magnetic-field facility with a low noise level at HFML-FELIX, Radboud University, the Netherlands. The magnetic field was applied using a wide-bore $\unit[38]{T}$ resistive magnet. The AC bias current was applied using a Keithley 6221 DC and AC current source at a frequency of $\unit[37]{Hz}$, and the harmonic voltages were detected using Stanford Research System SR860 $\unit[500]{kHz}$ DSP lock-in amplifiers.

The transport measurements in Device 1, 3 and 4 were performed in a Quantum Design PPMS DynaCool system. Stanford Research System SR830 $\unit[100]{kHz}$ DSP lock-in amplifiers were used both for applying the AC bias current at a frequency of $\unit[17]{Hz}$ for Device 1 and 4, and $\unit[433]{Hz}$ for Device 3, and for detecting the harmonic voltages.

\emph{Theoretical calculations:} The MBT system was modeled using a $k.p$ model which has been fitted to first-principles computations and allows for the consideration of the effects of a finite magnetic field.\cite{Li2024QuantumMnBi2Te4} Higher-order contributions to the longitudinal conductivity from Drude-like terms and the quantum metric were then calculated from the acquired dispersion relation. The theoretical calculations are described in detail in Supplementary Note 4.

\section*{Data availability}
The data supporting the findings of this study are available from the corresponding author upon a reasonable request.

\begin{acknowledgement}

The authors acknowledge financial support from KAW-WISE (Wallenberg Initiative Materials Science for Sustainability), Swedish Research Council VR project (No. 2025-03702), 2D TECH VINNOVA competence center (No. 2019-00068), FLAG-ERA project Magic Tune (2023-06210), Areas of Advance Nano, Energy, and Materials Science at Chalmers University of Technology. We acknowledge the help of staff at Quantum Device Physics and Nanofabrication laboratory in our department at Chalmers. Devices were fabricated at Nanofabrication laboratory, Myfab, MC2, Chalmers. We acknowledge the support of the HFML, member of the European Magnetic Field Laboratory (EMFL). C. M. Canali and S. Sattar acknowledge financial support from the Swedish Research Council (grant no: VR 2021-04622). The computations were enabled by resources provided by the National Academic Infrastructure for Supercomputing in Sweden (NAISS) partially funded by the Swedish Research Council through grant agreement no. 2022-06725 and the Centre for Scientific and Technical Computing at Lund University (LUNARC).

\end{acknowledgement}

\section{{\large Author information}}
\subsection{{\normalsize Contributions}}

L.S. fabricated the devices together with A.K., P.R. and E.R.. A.T. grew the material. L.S. characterized the devices together with P.R., M.B., E.R., A.K. and U.Z.. L.S., P.R. and S.P.D. conceived the idea and designed the experiments. S.S., A.T. and C.C. performed the theoretical calculations. L.S. analyzed and interpreted the experimental data, compiled the figures, and wrote the manuscript with inputs from all co-authors. S.P.D. supervised the research project.

\subsection{{\normalsize Competing interests}}
The authors declare no competing financial or non-financial interests.




\bibliography{references}

\end{document}